\documentclass[11pt,fleqn,twoside]{article}

\usepackage{amsmath}
\usepackage{amsthm}
\usepackage{amssymb}
\makeatletter
\newcommand{\prava}[1]{\small\it
\begin{flushleft}
Copyright \copyright \ 2000 by  #1
\end{flushleft}}

\newcommand{\name}[1]{\begin{flushleft}
                       \LARGE \bf #1
                       \end{flushleft}\vspace{-3mm}}

\newcommand{\Author}[1]{\begin{flushleft}
                       \it #1 \end{flushleft}}

\newcommand{\Adress}[1]{\begin{flushleft}
                       \it #1 \end{flushleft}}

\newcommand{\Date}[1]{\begin{flushleft}
                      \small  \it #1 \end{flushleft}}

\newcommand{\ehkol}{Author \ name}
\newcommand{\ohkol}{Article \ name}

\renewcommand{\@evenhead}{
\hspace*{-3pt}\raisebox{-15pt}[\headheight][0pt]{\vbox{\hbox to \textwidth 
{\thepage \hfil \ehkol}\vskip4pt \hrule}}}
\renewcommand{\@oddhead}{
\hspace*{-3pt}\raisebox{-15pt}[\headheight][0pt]{\vbox{\hbox to \textwidth 
{\ohkol \hfil \thepage}\vskip4pt\hrule}}}
\renewcommand{\@evenfoot}{}
\renewcommand{\@oddfoot}{}

\long\def\@makecaption#1#2{%
  \vskip\abovecaptionskip
  \sbox\@tempboxa{\small \textbf{#1.}\ \ #2}%
  \ifdim \wd\@tempboxa >\hsize
    {\small \textbf{#1.}\ \ #2}\par
  \else
    \global \@minipagefalse
    \hb@xt@\hsize{\hfil\box\@tempboxa\hfil}%
  \fi
  \vskip\belowcaptionskip}

\def\numberwithin#1#2{\@ifundefined{c@#1}{\@nocounterr{#1}}{%
  \@ifundefined{c@#2}{\@nocnterr{#2}}{%
  \@addtoreset{#1}{#2}%
  \toks@\@xp\@xp\@xp{\csname the#1\endcsname}%
  \@xp\xdef\csname the#1\endcsname
    {\@xp\@nx\csname the#2\endcsname
     .\the\toks@}}}}

\renewenvironment{proof}[1][\proofname]{\par
  \normalfont
  \topsep6\p@\@plus6\p@ \trivlist
  \item[\hskip\labelsep\textbf{%
    #1}\@addpunct{.}]\ignorespaces
}{%
  \qed\endtrivlist
}

\newcommand{\resetfootnoterule} {
  \renewcommand\footnoterule{%
  \kern-3\p@
  \hrule\@width.4\columnwidth
  \kern2.6\p@}
}

\makeatother

\numberwithin{equation}{section}

\allowdisplaybreaks[3]
\def\Real   {\mathbb{R}}
\def\Complex{\mathbb{C}}
\def\Field  {\mathbb{F}}
\def\phi{\varphi}
\def\eps{\varepsilon}
\def\la {\lambda}
\def\pa {\partial}
\def\ti {\tilde}
\def\const{\mathop{\rm const}\nolimits}
\def\tr   {\mathop{\rm tr}   \nolimits}
\def\ot   {\mathop{\otimes}}
\def\12{{1\over2}}
\def\<{\langle}
\def\>{\rangle}
\def\LRA{\ \Leftrightarrow\ }
\def\ba{\begin{array}}              \def\ea{\end{array}}
\def\bean{\begin{eqnarray*}}          \def\eean{\end{eqnarray*}}
\newtheorem{theorem}{Theorem}
\newtheorem{lemma}[theorem]{Lemma}
\newtheorem{prop}[theorem]{Proposition}
\newtheorem{definition}{Definition}

\makeatletter
\DeclareRobustCommand{\primfrac}[1]{%
  \PackageWarning{amsmath}{%
Foreign command \@backslashchar#1; %
\protect\frac\space or \protect\genfrac\space should be used instead%
  }
  \global\@xp\let\csname#1\@xp\endcsname\csname @@#1\endcsname
  \csname#1\endcsname
}
\makeatother

\begin{document}

\thispagestyle{empty}
\renewcommand{\ehkol}{V.E.\ Adler}
\renewcommand{\ohkol}{B\"acklund Transformations
for the Relativistic Lattices}

\begin{flushleft}
\footnotesize \sf
Journal of Nonlinear Mathematical Physics \qquad 2000, V.7, N~1,
\pageref{adler_fp}--\pageref{adler_lp}.
\hfill {\sc Article}
\end{flushleft}

\vspace{-5mm}

\renewcommand{\footnoterule}{}
{\renewcommand{\thefootnote}{}
 \footnotetext{\prava{V.E.\ Adler}}}

\name{On the Structure of the B\"acklund Transformations
for the Relativistic Lattices}\label{adler_fp}

\Author{Vsevolod E.\ ADLER}

\Adress{Ufa Institute of Mathematics, 
        112 Chernyshevsky str., 
        450077 Ufa, Russia\\
        e-mail: adler@imat.rb.ru}

\Date{Received July 22, 1999; Revised September 5, 1999; Accepted
September 9, 1999}

\begin{abstract}
\noindent
The B\"acklund transformations for the relativistic lattices of the Toda
type and their discrete analogues can be obtained as the composition of two
duality transformations.  The condition of invariance under this composition
allows to distinguish effectively the integrable cases.  Iterations of the
B\"acklund transformations can be described in the terms of nonrelativistic
lattices of the Toda type.  Several multifield generalizations are
presented.
\end{abstract}

\section{Introduction}

In this paper we consider four classes of difference and
differential-difference equations, which contain, in particular, the Toda
lattice introduced by Toda \cite{Toda}, the relativistic Toda lattice
introduced by Ruijsenaars \cite{Rui86, Rui90}, their generalizations
introduced by Yamilov \cite{Y93} and Suris \cite{Sur97a,Sur97b,Sur97c} and
their discretizations introduced by Hirota \cite{Hir} and Suris
\cite{Sur96,Sur90,Sur95}.  The systematic account is given of the method
proposed in the papers \cite{AS2,A} for studying of these classes of
equations, and some classification results are presented.  The main idea of
the method can be described in a few words.

For the given lattice equation $e[q]=0$ we define, in a certain way, the
pair of transformations $T_+:q\to Q$ and $T_-:q\to\ti Q.$  In general,
variables $Q$ and $\ti Q$ satisfy the distinct equations.  We say that the
lattice $e[q]=0$ admits the duality transformations $T_\pm$ if both
variables $Q,$ $\ti Q$ satisfy the same equation $E[Q]=0$ which is called
the dual equation. Obviously, the composition of the duality transformations
can be used for reproducing of solutions.  Denoting iterations of the
transformations $T_l=T^{-1}_-T_+$ and $T^+_l=T_+T^{-1}_-$ by superscript $l$
one obtains the commutative diagram displayed on the Figure~1.

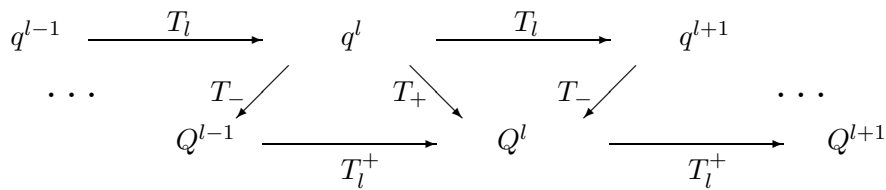
\begin{figure}[!ht]
\setlength{\unitlength}{0.06em}
\centering
\begin{picture}(550,120)(-50,-20)
 \put(  0,27){\LARGE$\dots$}
 \put(420,27){\LARGE$\dots$}
 \put(-20,57){$q^{l-1}$}
 \put(170,57){$q^l$}
 \put(365,57){$q^{l+1}$}
 \put( 75,-3){$Q^{l-1}$}
 \put(260,-3){$Q^l$}
 \put(450,-3){$Q^{l+1}$}
 \multiput( 70, 65)(200,0){2}{$T_l$}
 \multiput(170,-20)(200,0){2}{$T^+_l$}
 \multiput( 25,60)(200,0){2}{\vector( 1, 0){100}}
 \multiput(125, 0)(200,0){2}{\vector( 1, 0){100}}
 \multiput(140,45)(200,0){2}{\vector(-1,-1){ 30}}
 \multiput( 95,25)(200,0){2}{$T_-$}
 \put(200,25){$T_+$}
 \put(210,45){\vector(1,-1){30}}
\end{picture}
\caption{B\"acklund transformation as the composition of
         duality transformations.}
\end{figure}

In other words, transformations $T_l$ and $T^+_l$ define the B\"acklund
transformations for the given lattice and its dual.  This explains the
connection between duality transformations and integrability since existence
of the B\"acklund transformation is an indispensable feature of any
integrable system.  Quite analogously, the B\"acklund transformations for
the KdV and mKdV equations are obtained by the composition of two slightly
different Miura maps and the Schlesinger transformation for PII is
constructed from two substitutions into PXXXIV.

The requirement that the variables $Q$ and $\ti Q$ satisfy the same equation
is stringent enough and allows to distinguish effectively the integrable
relativistic lattices. However, it does not work for the subclass of
nonrelativistic lattices, which are characterized by the property $T_+=T_-.$
In contrast to the relativistic case, the duality transformation is now
irrelevant to integrability and cannot be used for classification.
Nevertheless, this subclass does not require special treatment since it
arises as a by-product of already obtained results.  Namely, it turns out
that the iterations of the transformation $T_l$ for the relativistic lattice
admitting duality transformation are described by some lattice of the Toda
type and the same is true for their discrete analogues.

The above scheme is applied to the discrete relativistic lattices in the
Section \ref{s:drtl}.  In the Section \ref{s:ddt} we introduce the notions
of the duality transformation and the dual equation, and prove that the dual
equation belongs to the same class.  In the Section \ref{s:dtl} we study the
B\"acklund transformation $T_l$ in more details and prove that it is
equivalent to some discrete lattice of the Toda type.  The classification of
integrable equations is performed in the Section \ref{s:dlist}, several
multifield generalizations are presented in the Section \ref{s:dmult}.  The
Section \ref{s:rtl} devoted to the lattices of the relativistic Toda type
is, in fact, exact continuous double of the Section \ref{s:drtl}. Joint of
both discrete and continuous theories is performed in the Section
\ref{s:nsp}.

\section{The lattices of the discrete relativistic Toda type}\label{s:drtl}

This section deals with the difference equations on the real or complex
variable $q$ defined on the two-dimensional integer lattice.  As a rule, we
use abridged notation $q=q_{mn},$ $q_{ij}=q_{m+i,n+j}.$  The shift operators
on the first and second subscripts will be denoted $T_m$ and $T_n$
respectively.  Next, it is convenient to introduce notation for the
differences (see Figure~2)
\[
    x=q-q_{-1,0}, \quad  y=q-q_{0,-1}, \quad  z=q-q_{-1,-1}.
\]
Obviously, one of these differences can be expressed through the other two,
e.g. $z=x+y_{-1,0}=y+x_{0,-1},$ and the following identity is valid
\begin{equation}\label{xy}
    (T_m-1)y_{0,1} = (T_n-1)x_{1,0}.
\end{equation}
The uppercase letters $Q,X,Y,Z$ (reserved for the dual variable and its
differences) are used in quite similar manner.

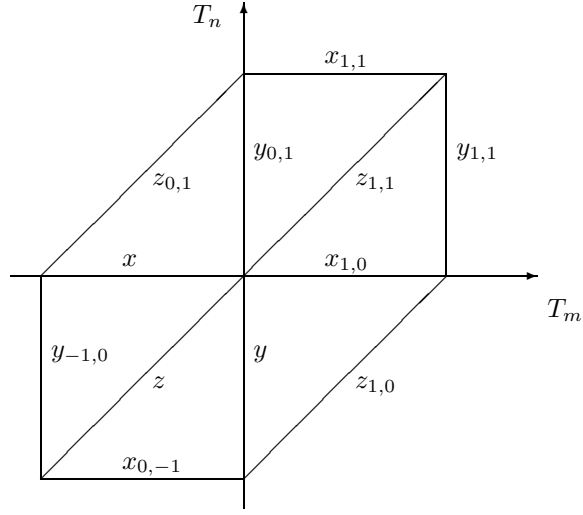
\begin{figure}[t]
\setlength{\unitlength}{0.07em} \small
\centering
\begin{picture}(290,270)(-120,-120)
 \put(-115,0){\vector(1,0){260}} \put(150,-20){$T_m$}
 \put(0,-115){\vector(0,1){250}} \put(-25,125){$T_n$}
 \put(-100,-100){\line(1,1){200}}
 \multiput(-100,-100)(200,100){2}{\line(0,1){100}}
 \multiput(-100,-100)(100,200){2}{\line(1,0){100}}
 \multiput(-100,0)(100,-100){2}{\line(1,1){100}}
 \put(-60,  5){$x$}        \put( 40,  5){$x_{1,0}$}
 \put(-60,-95){$x_{0,-1}$} \put( 40,105){$x_{1,1}$}
 \put(-95,-40){$y_{-1,0}$} \put(  5,-40){$y$}
 \put(  5, 60){$y_{0,1}$}  \put(105, 60){$y_{1,1}$}
 \put(-45,-55){$z$}        \put( 55,-55){$z_{1,0}$}
 \put(-45, 45){$z_{0,1}$}  \put( 55, 45){$z_{1,1}$}
\end{picture}
\normalsize
\caption{Pattern for the discrete relativistic lattice.}
\end{figure}

\subsection{Duality transformations}\label{s:ddt}

The lattices of the discrete relativistic Toda type
\begin{equation}\label{drtl}
    (T_m-1)f(x) + (T_n-1)g(y) + (T_mT_n-1)h(z) = 0
\end{equation}
are the Euler equations for the Lagrangians of the form
\begin{equation}\label{dL}
    {\cal L} = \sum_{m,n}(a(x) + b(y) + c(z))
\end{equation}
where $f=a',$ $g=b',$ $h=c'.$  In this section we assume that $f'g'h'\ne0$
in order to eliminate the case of the nonrelativistic lattices, so that
equation (\ref{drtl}) involves 7 nodes of the lattice as shown on the
Figure~2.  It is clear that the roles of the shifts $T_m,T_n$ and
$T^{-1}_mT^{-1}_n$ in (\ref{drtl}) are equal (actually, one can assume them
as the generators of the regular hexagonal lattice).

Equation (\ref{drtl}) can be rewritten in two equivalent forms of the
momentum conservation law:
\begin{gather*}
         (T_m-1)(f(x)+h(z_{0,1})) + (T_n-1)(g(y)+h(z)) = 0 \\
    \LRA (T_m-1)(f(x)+h(z)) + (T_n-1)(g(y)+h(z_{1,0})) = 0.
\end{gather*}
This allows to introduce the pair of transformations acting on the
differences:
\begin{equation}\label{dTpm}
    T_+:\ (X,Y_{-1,0})= T(x,y_{0,1}),\qquad
    T_-:\ (\ti X_{0,-1},\ti Y) = T(x_{1,0},y)
\end{equation}
where the mapping $T:(x,y)\to (X,Y)$ is given by the formulae
\begin{equation}\label{dT}
    X = g(y)+h(x+y), \quad Y = -f(x)-h(x+y).
\end{equation}

Obviously, the variables $Q$ and $\ti Q$ corresponding to $X,Y$ and $\ti
X,\ti Y$ are defined only up to the addition of an arbitrary constant.
However, the equations for these variables contain only differences and can
be derived by solving (\ref{dTpm}), (\ref{dT}) with respect to $x,y$ and
using the identity (\ref{xy}). Generally, these equations are distinct from
each other.

\begin{definition}\label{d:ddt}
The lattice (\ref{drtl}) admits the duality transformations (\ref{dTpm}),
(\ref{dT}) if the mapping $T$ is invertible and both variables $Q,$ $\ti Q$
satisfy the same lattice which is called dual to (\ref{drtl}).
\end{definition}

The following Theorem characterizes equations admitting duality
transformations in terms of the mapping $T.$  It also demonstrates that the
duality transformations do not lead out off the class (\ref{drtl}).  The
immediate corollary is that the equation which is dual to the dual equation
coincides with the original one.

\begin{theorem}\label{th:ddt}
Equation (\ref{drtl}) admits duality transformations if and only if the
inverse of (\ref{dT}) is of the form
\begin{equation}\label{dinvT}
    x = G(Y)+H(X+Y), \quad  y = -F(X)-H(X+Y).
\end{equation}
In this case the dual equation is of the form
\begin{equation}\label{DRTL}
    (T_m-1)F(X) + (T_n-1)G(Y) + (T_mT_n-1)H(Z) = 0.
\end{equation}
\end{theorem}
\begin{proof}
Let $T^{-1}$ be of the stated form, then one can easily check that the
elimination of $x,y$ by means of the identity (\ref{xy}) brings to equation
(\ref{DRTL}) for both transformations (\ref{dTpm}).

Conversely, assume that equations for $Q$ and $\ti Q$ coincide.  Let $T^{-1}$
be of the form
\[
    x=\Phi(X,Y), \quad y=\Psi(X,Y)
\]
then inverses of the transformations (\ref{dTpm}) are given by the formulae
(the tilde in the second one is omitted)
\begin{align*}
    T^{-1}_+:&\quad x=\Phi(X,Y_{-1,0}), \quad y_{0,1}=\Psi(X,Y_{-1,0}), \\
    T^{-1}_-:&\quad x_{1,0}=\Phi(X_{0,-1},Y), \quad y=\Psi(X_{0,-1},Y).
\end{align*}
The identity (\ref{xy}) yields the equations
\begin{gather*}
    \Phi(X_{1,1},Y_{0,1})-\Phi(X_{1,0},Y)-\Psi(X_{1,0},Y)+\Psi(X,Y_{-1,0})=0,
    \\[1ex]
    \Phi(X,Y_{0,1})-\Phi(X_{0,-1},Y)-\Psi(X_{1,0},Y_{1,1})+\Psi(X,Y_{0,1})=0
\end{gather*}
which must be equivalent to each other.  In order to compare these equations,
rewrite the last one in the form
\[
    \Phi(X,Y_{0,1}) - \Phi(X+Y_{-1,0}-Y,Y)
    - \Psi(X_{1,0},X_{1,1}+Y_{0,1}-X_{1,0}) + \Psi(X,Y_{0,1})=0,
\]
so that both equations contain the variables $X_{1,1},$ $X_{1,0},$ $X,$
$Y_{0,1},$ $Y,$ $Y_{-1,0}.$  Now let us consider $X_{1,1}$ as function on
the rest variables from this set, then
\begin{align*}
    {\pa X_{1,1}\over \pa Y_{0,1}}
    &= -{\pa_{Y_{0,1}}\Phi(X_{1,1},Y_{0,1}) \over
        \pa_{X_{1,1}}\Phi(X_{1,1},Y_{0,1})} \\
    &= {\pa_{Y_{0,1}}[\Phi(X,Y_{0,1})-\Psi(X_{1,0},X_{1,1}+Y_{0,1}-X_{1,0})
                    +\Psi(X,Y_{0,1})] \over
        \pa_{X_{1,1}}\Psi(X_{1,0},X_{1,1}+Y_{0,1}-X_{1,0})}.
\end{align*}
The last equality must be satisfied identically.  Differentiating it with
respect to $X$ and $X_{1,0}$ yields
\[
    \pa_X\pa_{Y_{0,1}}(\Phi(X,Y_{0,1})+\Psi(X,Y_{0,1}))=0,\quad
    \pa_{X_{1,0}}\pa_{Y_{0,1}}\Psi(X_{1,0},X_{1,1}+Y_{0,1}-X_{1,0})=0
\]
and hence
\begin{equation}\label{phipsi}
    \Phi(X,Y)+\Psi(X,Y)=G(Y)-F(X),\quad \Psi(X,Y)=K(X)-H(X+Y).
\end{equation}
Analogously, differentiating the relation
\begin{align*}
    {\pa Y_{-1,0}\over\pa X}
    &= -{\pa_X\Psi(X,Y_{-1,0}) \over \pa_{Y_{-1,0}}\Psi(X,Y_{-1,0})} \\
    &= {\pa_X[\Phi(X,Y_{0,1})-\Phi(X+Y_{-1,0}-Y,Y)+\Psi(X,Y_{0,1})] \over
       \pa_{Y_{-1,0}}\Phi(X+Y_{-1,0}-Y,Y)}
\end{align*}
with respect to $Y$ yields
\[
    \pa_X\pa_Y\Phi(X+Y_{-1,0}-Y,Y)=0\ \Rightarrow\ \Phi(X,Y)=L(Y)+M(X+Y).
\]
Comparing with (\ref{phipsi}) completes the proof.
\end{proof}

\subsection{Classification theorem}\label{s:dlist}

The Definition \ref{d:ddt} turns out to be severe enough and allows to
obtain the finite list of integrable equations (\ref{drtl}).  In virtue of
the Theorem \ref{th:ddt} it is sufficient to find all functions $f,g,h$ such
that inverse of the transformation (\ref{dT}) is given by (\ref{dinvT}).
This means that the Jacobian $\Delta=f'g'+g'h'+h'f'$ of the map (\ref{dT})
must be nonzero and the following identities must hold
\[
    x_{XY}+y_{XY}=0,\quad  x_{XY}=x_{XX},\quad  y_{XY}=y_{YY}.
\]
These three relations are equivalent.  Indeed, the Jacobi matrix is
\[
    \left(\ba{cc} x_X & y_X \\ x_Y & y_Y \ea\right) =
    {1\over\Delta}
    \left(\ba{cc} -h' & f'+h' \\ -g'-h' & h' \ea\right)
\]
that is $x_X=-y_Y.$  Straightforward computation proves that functions
$f(x),$ $g(y),$ $h(x+y)$ must satisfy the equation
\[
    (g'+h'){f''\over f'} + (f'+h'){g''\over g'} = (f'+g'){h''\over h'}.
\]
The designations $f'=1/u,$ $g'=1/v,$ $h'=1/w$ rewrite it in more convenient
form
\begin{equation}\label{uvw}
    [v(y)+w(x+y)]u'(x) + [u(x)+w(x+y)]v'(y) = [u(x)+v(y)]w'(x+y).
\end{equation}
The classification problem is reduced to solving of this functional
equation. Of course, functions $u,v,w$ can be multiplied by an arbitrary
constant and the linear transformation
\[
    \ti x=c(x-x_0), \quad \ti y=c(y-y_0) \LRA
    \ti q_{m,n}=c(q_{m,n}-mx_0-ny_0-\const)
\]
can be applied, as well as the permutation of the $x,y,z$ axes.

At first let us consider some degenerate cases.  Assume that two of three
functions are constant, say $u$ and $v.$  Then (\ref{uvw}) yields that
either $w$ is constant as well, or $u=-v$ and $w$ is arbitrary.  In the
first case the equation (\ref{drtl}) is linear and in the second one it is
of the form
\[
   \alpha(q_{1,0}+q_{-1,0}-q_{0,1}-q_{0,-1})+h(q_{1,1}-q)-h(q-q_{-1,-1}) = 0
\]
and admits ``integration'':
\[
   \alpha(q_{m+1,n}-q_{m,n+1})+h(q_{m+1,n+1}-q_{m,n}) = c_{m-n}.
\]
Other degenerate case corresponds to the vanishing of the Jacobian, what is
equivalent to $u+v+w=0$ and implies that all three functions are linear.  In
this case one can prove that equation (\ref{drtl}) can be reduced to the
equation on the variable $p=y_{0,1}/x.$  Further on we will not consider
these degenerate cases.

\begin{lemma}
The functions $u,v,w$ satisfy the equations
\begin{equation}\label{uvw'}
    (u')^2= \delta u^2+2\alpha u+\eps,\quad
    (v')^2= \delta v^2+2\beta  v+\eps,\quad
    (w')^2= \delta w^2+2\gamma w+\eps.
\end{equation}
\end{lemma}
\begin{proof}
At first prove that functions $u(x)$ and $v(y)$ satisfy the equation
\begin{equation}\label{uv}
    (u''-v'')(u+v)-(u')^2+(v')^2=k(u-v), \quad  k=\const.
\end{equation}
Let us eliminate $w$ from (\ref{uvw}).  Applying the operator $\pa_x-\pa_y$
one obtains the linear system on $w,w':$
\[
    \left(\ba{cc}
      u'+v'  & -u-v \\
     u''-v'' & v'-u'
    \ea\right) \left(\ba{c} w+u \\ w'  \ea\right) =
          (u-v)\left(\ba{c}  u' \\ u'' \ea\right).
\]
Its determinant $\Delta$ is exactly the left hand side of the equation
(\ref{uv}).  If it is identically zero then (\ref{uv}) is proved, otherwise
one finds
\[
    w+u= {u-v\over\Delta}(uu''-(u')^2+u''v+u'v'),\quad
     w'= {u-v\over\Delta}(u''v'+u'v'')
\]
and consequently $((u-v)/\Delta)_y(uu''-(u')^2+u''v+u'v')=0.$  Assume that
the expression in the second bracket vanishes.  If $u'\not\equiv0$ then
$v'=-u''v/u'+u'-uu''/u',$ $v''=-u''v'/u',$ but then, as one easily checks,
$\Delta=0.$  If $u'=0$ then $w+u=0,$ that is we come to the degenerate
solution excluded above.  Therefore $((u-v)/\Delta)_y=0.$  Due to the
symmetry between $u$ and $v$ one can prove analogously $((u-v)/\Delta)_x=0$
and obtain (\ref{uv}).

Further on, rewriting (\ref{uv}) in the form
\[
   \left({u'\over u+v}\right)_x = {v''+k\over u+v}-{(v')^2+2kv\over(u+v)^2},
\]
multiplying by $u'/(u+v)$ and integrating with respect to $x$ yield
\[
    (u')^2 =    \delta(y)(u+v)^2 - 2(v''+k)(u+v)+(v')^2+2kv.
\]
Replacing $u$ and $v$ one obtains
\[
    (v')^2 = \ti\delta(x)(u+v)^2 - 2(u''+k)(u+v)+(u')^2+2ku.
\]
Subtracting one equation from another and using (\ref{uv}) one obtains
$\ti\delta=\delta=\const.$  Summing and dividing by $u+v$ give
$u''+v''=\delta(u+v)-k.$  The separation of the variables yields
$(u')^2=\delta u^2+2\alpha u+\eps,$ $(v')^2=\delta v^2+2\beta v+\ti\eps,$
where $\alpha+\beta=-k,$ and substitution into (\ref{uv}) proves
$\eps=\ti\eps.$ The last of the equations (\ref{uvw'}) is obtained in virtue
of the symmetry of $x,y,z$ axes.
\end{proof}

It is clear that the solutions of equations (\ref{uvw'}) must satisfy also
some additional relations.  However their analysis is not in principle
difficult, and the direct examination of all solutions brings to the
following list.

\begin{theorem}\label{th:drtl}
The equations (\ref{drtl}) admitting duality transformations are exhausted,
up to the changes $\ti q_{m,n}=c(q_{m,n}-mx_0-ny_0)$ and permutations of
$x,y,z$ axes, by the following sets of the functions $f,g,h.$  In formulae
(A), (B), (C) the parameters are constrained by relation $\la+\mu+\nu=0,$
and in (I) by relation $\la\mu\nu=-1.$
\[\ba{lllll}
 (A) && f={\mu\over x}, &  g={\nu\over y}, &  h={\la\over z}, \\
 (B) && f=\mu\coth x,   &  g=\nu\coth y,   &  h=\la\coth z,   \\
 (C) && f=\12\log{x+\mu\over x-\mu}, &
        g=\12\log{y+\nu\over y-\nu}, &
        h=\12\log{z+\la\over z-\la},  \\
 (D) && f=\log x,          &  g=\log y,     &  h=\log(1-1/z),     \\
 (E) && f=-e^x-1,          &  g=e^{-y},     &  h={1\over 1+e^z},  \\
 (F) && f=\log(e^x-1),     & g=\log(e^y-1), &  h=-\log(e^z-1),    \\
 (G) && f=-\log(e^{-x}-1), & g=\log(e^y-1), &  h=-z,              \\
 (H) && f=\log(\la^{-1}(e^x+1)),   &
        g=\log(e^{-y}-1),          &
        h=\log{e^z+\la\over e^z+1}, \\
 (I) && f=\log{\mu e^x+1\over e^x+\mu}, &
        g=\log{\nu e^y+1\over e^y+\nu}, &
        h=\log{\la e^z+1\over e^z+\la}.
\ea\]
The duality transformations (\ref{dTpm}), (\ref{dT}) link together the
equations corresponding to solutions (B) and (C), (D) and (E), (F) and (G),
while the equations corresponding to solutions (A), (H) and (I) are
self-dual.
\qed
\end{theorem}

Notice, that the cases (A) and (B) are connected by the point transformation
$q=\exp(2\ti q).$  It is explained by the fact that the Lagrangian
$\sum(\mu\log x+\nu\log y+\la\log z)$ of the equation (\ref{drtl}), (A) is
invariant under the dilations $q\to Cq$ as well as under the shifts $q\to
q+C.$  On the other hand, the inversions $q\to q/(1-Cq)$ preserve the
Lagrangian as well but the change $q=1/\ti q$ which maps this group into the
shift group does not bring to a new equation.

\subsection{The lattices of the discrete Toda type}\label{s:dtl}

The method proposed in the Section \ref{s:ddt} does not work for the
lattices of the discrete Toda type which correspond to the case $h=0$ (this
is equivalent to $f'g'h'=0,$ without loss of generality). Indeed, in this
case transformations $T_+$ and $T_-$ coincide for arbitrary $f,g$ and
classification becomes impossible. However, we can dispense with it since
these lattices arise as a by-product of already obtained results for the
discrete relativistic lattices.  Namely, we will demonstrate, using only few
basic formulae from the Section \ref{s:ddt} and without any complicated
calculations, that the iterations of the B\"acklund transformation
$T_l=T^{-1}_-T_+$ are described by some discrete lattice of the Toda type.
Hence we automatically obtain some list of integrable lattices, see Theorem
\ref{th:dtl} below.  Probably, this list is exhaustive (cf.
\cite{Sur97a,Sur90,Sur95,MaSh}), but unfortunately I do not know any
classification results, like Yamilov's Theorem \ref{th:tl}, which can be
compared with this list.

Let us consider some lattice (\ref{drtl}) admitting duality transformations
and denote iterations of $T_l$ by superscript $l,$ in such a way that tilde
in the formula (\ref{dTpm}) corresponds to the value $l-1.$  It is possible
to rewrite equations (\ref{drtl}), (\ref{DRTL}) in terms of the mixed
variables $x,X.$  Indeed, one obtains directly from (\ref{dTpm}), (\ref{dT})
the relations
\[
    X_{0,-1}= g(y) + h(x_{0,-1}+y),\quad
    X^{-1}  = g(y_{0,1}) + h(x_{1,1}+y_{0,1})
\]
and therefore equation (\ref{drtl}) is equivalent to
\[
    (T_m-1)f(x) + X^{-1} - X_{0,-1} = 0.
\]
Analogously, in virtue of (\ref{dinvT}) equation (\ref{DRTL}) is equivalent
to
\[
    (T_m-1)F(X) + x_{1,1}-x^1_{1,0} = 0.
\]
``Integrating'' this equation with respect to $m$ (recall that
$x_{1,0}=(T_m-1)q$) one obtains $X=\phi((T_l-T_n)q+c)$ where function $\phi$
is inverse of $F.$  The constant $c$ does not depend on $m,$ but may depend
on $l,n,$ and it can be set to zero without lost of generality by means of
appropriate shift of the variables $q.$  Then eliminating $X$ from the
previous equation brings to the lattice of the discrete Toda type
\begin{equation}\label{xdtl}
    (T_m-1)f(x) - (T_lT^{-1}_n-1)\phi((1-T^{-1}_lT_n)q) = 0.
\end{equation}

So, we have already proved that the B\"acklund transformation for the
discrete relativistic lattice is governed by some discrete nonrelativistic
lattice.  However, the complete picture is even more rich: it turns out that
the 3-dimensional lattice generated by the shifts $T_l,$ $T_m,$ $T_n$
contains 3 instances of the discrete nonrelativistic lattices and 4
instances of the discrete relativistic lattices. In order to see this let us
remind that the roles of all shifts in equation (\ref{drtl}) are equal and
therefore we can repeat the above calculation starting from the other set of
mixed variables.  More precisely, rewriting equations (\ref{drtl}) and
(\ref{DRTL}) in terms of $y,Y$ or $z,Z$ one obtains the equations
\begin{gather*}
    (T_n-1)g(y) + Y^{-1}_{-1,0} - Y = 0, \quad
    (T_n-1)G(Y) + y_{0,1} - y^1_{1,1} = 0, \\
    (T_mT_n-1)h(z) + Z - Z^{-1} = 0, \quad
    (T_mT_n-1)H(Z) + z^1_{1,1} - z_{1,1} = 0
\end{gather*}
and further elimination of $Y$ and $Z$ yields equations
\begin{gather}
    (T_n-1)g(y) - (T_lT_m-1)\psi((1-T^{-1}_lT^{-1}_m)q) = 0, \label{ydtl} \\
    (T_mT_n-1)h(z) + (T_l-1)\eta((T^{-1}_l-1)q) = 0 \label{zdtl}
\end{gather}
where functions $\psi$ and $\eta$ are inverses of $G$ and $H$ respectively.

\begin{figure}[t]
\setlength{\unitlength}{0.075em}
\centering
\begin{picture}(490,350)(-235,-160)

 \put(0,100){\vector(0,1){80}} \put(10,170){$n$}          
 \multiput(-180,-140)(100,  0){2}{\line(0,1){200}}
 \multiput(  20,-140)( 80, 40){3}{\line(0,1){200}}
 \multiput(-100,-100)( 80, 40){2}              {
    \multiput(0,0)(100,0){2}                  {
       \multiput(0,0)(0,30){7}{\line(0,1){20}}}}

 \put(100,0){\vector(1,0){140}} \put(220,-15){$m$}        
 \multiput(-180,-140)(  0,100){2}{\line(1,0){200}}
 \multiput(-180,  60)( 80, 40){3}{\line(1,0){200}}
 \multiput(-100,-100)( 80,40){2}               {
    \multiput(0,0)(0,100){2}                  {
       \multiput(0,0)(30,0){7}{\line(1,0){20}}}}

 \put(-80,-40){\vector(-2,-1){140}} \put(-225,-90){$l$}   
 \multiput(-180, 60)(100,   0){2}{\line(2,1){160}}
 \multiput(  20, 60)(  0,-100){3}{\line(2,1){160}}
 \multiput(-180,-140)(100,0){2}                {
    \multiput(0,0)(0,100){2}                  {
       \multiput(0,0)(24,12){7}{\line(2,1){16}}}}

 \multiput(0,-100)(0,200){2}
  {\hspace{-0.4em}\vrule width0.8em height0.4em depth0.4em} 
 \multiput(14,-46)(-40,80){2}{\framebox(12,12){}}           
 \multiput(-100,0)(200,0){2}{\circle*{12}}                  
 \multiput(-80,-140)(160,280){2}{\circle{12}}               
 \multiput(-87,-45)(160,80){2}
  {\put(0,0){\line(1,0){14}}
   \put(0,0){\line(1,2){7}}
   \put(14,0){\line(-1,2){7}}}                              
 \font\symbol=msam10 \def\bltr{{\symbol \char'116}}
 \multiput(-100,-100)(200,200){2}
  {\put(-6,-4){\bltr} \put(-3.5,-4){\bltr}
   \put(-1,-4){\bltr} \put(-3.5,1.2){\bltr}}                
 \end{picture}
 \caption{Each of three discrete nonrelativistic lattices involves 2 pairs
          of the same shape (black and white).  Each of four discrete
          relativistic lattices involves 3 distinct pairs, number of
          white pairs is even (0 or 2).}
\end{figure}
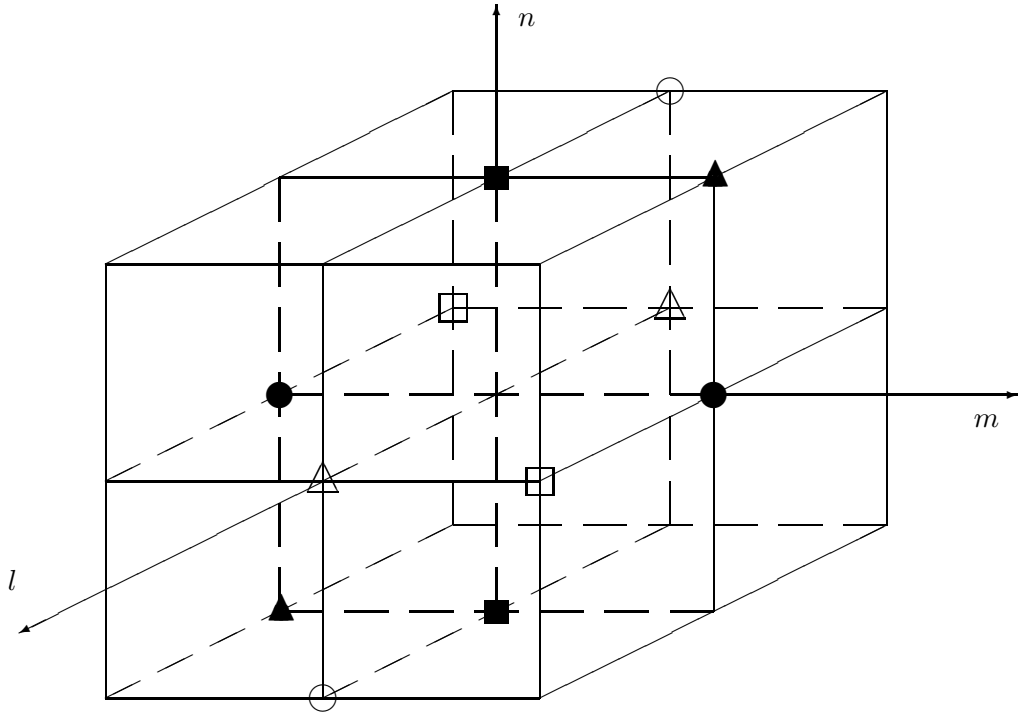

The placement of the variables involved in equations (\ref{xdtl}),
(\ref{ydtl}) and (\ref{zdtl}) is shown at the Figure~3.  At this picture the
variables $q$ involved in the original equation (\ref{drtl}) lie in the
plane $(mn)$ and are marked by black symbols, while the variables obtained
by means of the B\"acklund transformations are marked by the white ones.
Equation (\ref{xdtl}) constraints two black and two white circles (and, of
course, the centre of the cube), squares and triangles correspond to
(\ref{ydtl}) and (\ref{zdtl}).

Next, subtracting from the equation (\ref{drtl}) two of three equations
(\ref{xdtl}), (\ref{ydtl}) and (\ref{zdtl}) we obtain some discrete
relativistic lattice again, for example subtracting of (\ref{ydtl}),
(\ref{zdtl}) yields equation
\[
    (T_m-1)f(x) - (T_l-1)\eta((T^{-1}_l-1)q)
        + (T_lT_m-1)\psi((1-T^{-1}_lT^{-1}_m)q) = 0
\]
which involves variables in the plane $(lm).$  Two other choices bring to
the discrete relativistic lattices in the plane $(ln)$ and in the plane
spanned over the vectors $(0,1,1)$ and $(1,1,0).$

In conclusion of this section we present the list of the discrete lattices
of the Toda type which is obtained by direct examination of the equations
listed in the Theorem \ref{th:drtl}.
\begin{theorem}\label{th:dtl}
The equations (\ref{xdtl}), (\ref{ydtl}), (\ref{zdtl}) are equivalent, up to
the renaming of the shifts and the linear changes
$\ti q_{m,n}=c(q_{m,n}-mx_0-ny_0),$ to the lattices
\begin{gather*}
    (T_m-1){1\over x}     = (T_n-1){1\over y},     \\
    (T_m-1)e^x            = (T_n-1)e^y,            \\
    (T_m-1){1\over e^x-1} = (T_n-1){1\over e^y-1}, \\
    (T_m-1)\log x         = (T_n-1)\log y,         \\
    (T_m-1)\log(1-1/x)    = (T_n-1)\log(1-1/y),    \\
    (T_m-1)\log(e^x-1)    = (T_n-1)\log(e^y-1),    \\
    (T_m-1)x              = (T_n-1)\log(e^y-1),    \\
    (T_m-1)\log\left({e^x-\la\over e^x-1}\right) =
    (T_n-1)\log\left({e^y-\la\over e^y-1}\right).
\end{gather*}
\end{theorem}

\section{The lattices of the relativistic Toda type}\label{s:rtl}

Now let us consider the differential-difference equations on the variable
$q_n(t).$  As before we will omit the subscript $n:$ $q=q_n,$ $q_i=q_{n+i}$
and use uppercase letters for the dual variables.  Instead of differences
$x,y,z$ we consider the quantities
\[
    p=\dot q, \quad  y=q-q_{-1}
\]
which satisfy identity
\begin{equation}\label{py}
    D_ty_1 = (T_n-1)p.
\end{equation}

\subsection{Duality transformations}\label{s:dt}

The lattices of the relativistic Toda type
\begin{equation}\label{rtl}
    \dot p= r(p)(h(y_1)p_1-h(y)p_{-1}+g(y_1)-g(y)),  \quad
    \dot y= p-p_{-1}
\end{equation}
are the Euler equations for the Lagrangians of the form
\begin{equation}\label{L}
    {\cal L}= \int dt\sum_n(a(p)-b(y)-c(y)p),
\end{equation}
where $r=1/a'',$ $g=b',$ $h=c'.$  In this section we assume that the
nondegeneracy condition $h\ne0,$ $|g'|+|h'|\ne0$ is fulfilled.

The term $c(y)p$ of the Lagrangian is equivalent, up to the total
divergence, to $c(y_1)p$ and this results in two equivalent forms of the
momentum conservation law:
\begin{gather*}
    D_t(a'(p)-c(y_1)) = (T_n-1)(h(y)p_{-1}+g(y)) \\
    \LRA D_t(a'(p)-c(y)) = (T_n-1)(h(y)p+g(y)).
\end{gather*}
This allows to define the pair of transformations
\begin{equation}\label{Tpm}
    T_+:\ (P,Y) = T(p,y_1), \quad  T_-:\ (\ti P_{-1},\ti Y) = T(p,y)
\end{equation}
where mapping $T:(p,y)\to (P,Y)$ is defined by the formula
\begin{equation}\label{T}
    P = h(y)p+g(y), \quad Y = a'(p)-c(y).
\end{equation}

Notice, that the lattices (\ref{rtl}) are the continuous limit of the
discrete relativistic lattices (\ref{drtl}).  Indeed, let us consider the
family of the functionals of the form (\ref{dL})
\[
    {\cal L}_\eps = \sum_{m,n}(\eps a(x_{m,n}/\eps) - \eps b(y_{m,n})
                               + k(y_{m,n}) - k(z_{m,n})),
\]
and let $k'=c,$ $q_{m,n}=q_n(t),$ $t=m\eps.$  It is easy to see that passage
to the limit $\eps\to0$ brings exactly to the Lagrangian (\ref{L}).
Transformations (\ref{Tpm}) also are obtained from (\ref{dTpm}) by this
limit.

\begin{definition} \label{d:dt}
The lattice (\ref{rtl}) admits the duality transformations (\ref{Tpm}),
(\ref{T}) if the mapping $T$ is invertible and both variables $Q,$ $\ti Q$
satisfy the same lattice which is called dual to (\ref{rtl}).
\end{definition}

Notice that functions $a,b,c$ are defined up to some linear transformations
and therefore mapping (\ref{T}) is not uniquely defined.  However, it is easy
to see that this arbitrariness corresponds to the linear changes of the
variables $P,Y$ and all lattices dual to the given equation (\ref{rtl}) are
equivalent under the changes $Q_n\to\alpha Q_n+\beta n+\gamma t.$

\begin{theorem} \label{th:dt}
The lattice (\ref{rtl}) admits duality transformations if and only if the
inverse of the mapping (\ref{T}) is of the form
\begin{equation}\label{invT}
    p= H(Y)P+G(Y), \quad y= A'(P)-C(Y), \quad H=C'.
\end{equation}
In this case the dual equation is of the form ($R=1/A''$)
\begin{equation}\label{RTL}
    \dot P= R(P)(H(Y_1)P_1-H(Y)P_{-1}+G(Y_1)-G(Y)), \quad \dot Y= P-P_{-1}.
\end{equation}
\end{theorem}
\begin{proof}
Let $T^{-1}$ be of the form (\ref{invT}).  Then elimination of $p,y$ from
(\ref{Tpm}) by means of identity (\ref{py}) brings to equation (\ref{RTL})
in both cases.

In order to prove inverse statement we need to compare equations on the
variables $Q$ and $\ti Q.$  Straightforward calculation proves that the
variables $P,Y$ satisfy the equations of the form
\begin{equation}\label{mixed}
    \dot P= r(p)(h(y_1)(P-P_{-1})+\Delta(p,y_1)(p_1-p)),\quad
    \dot Y= P-P_{-1},
\end{equation}
where $\Delta(p,y)=h^2(y)+{1\over r(p)}(h'(y)p+g'(y))$ and $p_1,p,y_1$ have
to be expressed in terms of $P_1,P,Y_1,Y.$  Therefore, the dual equation
must be linear in $P_{-1}.$  Analogously, considering equation on $\ti Q$
one proves that the dual equation must be linear in $P_1.$ Since in
(\ref{mixed}) only $p_1$ depends on $P_1,$ hence mapping $T$ must satisfy
the condition $\pa^2p/\pa P^2=0.$  This proves the first formula in
(\ref{invT}).  The second one follows from the property $\pa y/\pa Y=-\pa
p/\pa P$ which is evident from the structure of the Jacobi matrix
\begin{equation}\label{J}
    \left(\ba{cc}
      \pa p/\pa P & \pa y/\pa P \\
      \pa p/\pa Y & \pa y/\pa Y
    \ea\right) = {1\over\Delta(p,y)}
    \left(\ba{cc}
              h(y)     & 1/r(p) \\
          h'(y)p+g'(y) & -h(y)
    \ea\right)
\end{equation}
where $\Delta\ne0$ by Definition \ref{d:dt}.
\end{proof}

As in the discrete case we see that the original lattice is dual for its
dual and the composition $T^{-1}_-T_+$ defines the B\"acklund transformation
for (\ref{rtl}) (see Figure~2). We will study this B\"acklund transformation
in the Section \ref{s:tl}.

\subsection{Classification theorem}\label{s:list}

\begin{theorem} \label{th:rgh}
The lattices (\ref{rtl}) admitting the duality transformations are
characterized by the following equations for the coefficients:
\begin{equation}\label{rgh}
\begin{split}
  & r=r_2p^2+r_1p+r_0,          \\
  & g'= r_2g^2+R_1g+R_0-r_0h^2, \quad
    h'= 2r_2gh-r_1h^2+R_1h.
\end{split}
\end{equation}
The coefficients of the dual lattice satisfy equations
\begin{equation}\label{RGH}
\begin{split}
  & R=r_2P^2+R_1P+R_0,          \\
  & G'= r_2G^2+r_1G+r_0-R_0H^2, \quad
    H'= 2r_2GH-R_1H^2+r_1H.
\end{split}
\end{equation}
\end{theorem}
\begin{proof}
\ Straightforward calculation proves that necessary and sufficient condition
\linebreak[4]%
\mbox{$\pa^2p/\pa P^2=0$} is equivalent to the following relation involving the
functions $r(p),g(y),h(y):$
\begin{equation}\label{rgh2}
    h(h''p+g'') - h'(h'p+g') + 2rh^2h' - r'h^2(h'p+g') = 0.
\end{equation}
The second derivative of (\ref{rgh2}) with respect to $p$ is
$r'''(h'p+g')=0.$  By the condition of nondegeneracy, this implies that $r$ is
a polynomial and its degree is less than 3.  Dividing (\ref{rgh2}) by $h^2$
and integrating the result with respect to $y$ one obtains
\[
    h'p+g' + 2rh^2 - r'h(hp+g) = h(\alpha+R_1p)
\]
where $\alpha$ and $R_1$ are some constants.  Let $r=r_2p^2+r_1p+r_0,$ then
collecting the coefficients of $p$ in this relation gives the system
\[
    g'= \alpha h + r_1gh - 2r_0h^2, \quad  h'= 2r_2gh - r_1h^2 + R_1h
\]
which is equivalent to (\ref{rgh}) modulo common first integral
\begin{equation}\label{alpha}
    r_2g^2+R_1g+R_0-r_1gh+r_0h^2 = \alpha h.
\end{equation}

The formula for $R$ can be easily obtained from the relation
$R=\Delta(p,y)r.$ Thus, equations for $g$ and $h$ are uniquely defined by
the coefficients of the polynomials $r$ and $R.$  Since due to the Theorem
\ref{th:dt} the duality relation is symmetric, hence equations for $G$ and
$H$ can be written automatically.
\end{proof}

\paragraph{Remark.}
The question about value of the first integral for the system (\ref{RGH}) is
a bit more complicated.  However, this value can be calculated by sequential
elimination of $g,h$ and $p$ from the formula (\ref{alpha}) using relations
$g=P-hp,$ $h=RH/r$ and $p=HP+G.$  This brings to relation
\[
    r_2G^2+r_1G+r_0-R_1GH+R_0H^2 = \alpha H,
\]
that is, the value of the integration constant $\alpha$ for the dual lattice
is the same.  Notice, that actually role of this constant is very important
since it may depend on $n.$  This brings to the integrable lattices
containing an arbitrary parameter in each node \cite{AS3} (it plays role of
the discrete spectrum when constructing solitons).  For sake of simplicity
we will not consider this generalization.

By use of the first integral (\ref{alpha}) the system (\ref{rgh}) is reduced
to the equation
\begin{equation}\label{h}
   (h')^2= (r^2_1-4r_2r_0)h^4 +(4\alpha r_2-2R_1r_1)h^3 +(R^2_1-4r_2R_0)h^2
\end{equation}
which is solved in elementary functions.  Consideration of all the possible
choices of parameters and the branches of solutions brings to the following
result.

\begin{theorem}\label{th:rtl}
Up to the transformations $q_n\to\alpha q_n+\beta t+\gamma n,$ $t\to\delta
t,$ the nondegenerate lattices (\ref{rtl}) admitting the duality
transformations are exhausted by the list $(\dot y=p-p_{-1}):$
\bean
 (a)     && \dot p= p_1e^{y_1}-p_{-1}e^y-e^{2y_1}+e^{2y},\\
 (b)     && \dot p= p\left({p_1\over y_1}-{p_{-1}\over y}+y_1-y \right), \\
 (c_{\mu,\nu})
         && \dot p= p\left( {p_1   \over1+\mu e^{-y_1}}
                           -{p_{-1}\over1+\mu e^{-y}}
                           +\nu(e^{y_1} - e^y) \right), \\
 (d)     && \dot p= p(p+1)\left({p_1\over y_1}-{p_{-1}\over y}\right), \\
 (e_\mu) && \dot p= p(p-\mu)\left( {p_1   \over\mu+e^{y_1}}
                                  -{p_{-1}\over\mu+e^y} \right),\\
 (f_\mu) && \dot p= (p^2+\mu)\left( {p_1-y_1 \over\mu+y^2_1}
                                   -{p_{-1}-y\over\mu+y^2} \right),\\
 (g_\mu) && \dot p= {1\over2}(p^2+1-\mu^2)
                     \left( {p_1-\sinh y_1 \over\mu+\cosh y_1}
                           -{p_{-1}-\sinh y\over\mu+\cosh y}\right).
\eean
The duality transformations (\ref{Tpm}), (\ref{T}) link together equations
(a) and (b), (d) and (e$_0$), (f$_\mu$) for $\mu\ne0$ and (g$_{\pm1}$),
while the rest equations are self-dual.  \qed
\end{theorem}

\subsection{The lattices of the Toda type}\label{s:tl}

As in the discrete case, the subclass of the Toda type lattices
\begin{equation}\label{tl}
    \dot p= r(p)(f(y_1)-f(y)),  \quad  \dot y= p-p_{-1}
\end{equation}
must be considered separately, since transformations $T_+$ and $T_-$
coincide.  It should be mentioned that in this case the duality
transformation was introduced by Toda \cite{Toda}. It is given by the
formula
\[
    P= f(y_1), \quad Y= a'(p), \quad a''=1/r
\]
and the coefficients of the dual lattice are defined by the formulae
$f'=R(f),$ $F(a'(p))=p.$  For example, the Toda lattice $\ddot q=
\exp(q_1-q)-\exp(q-q_{-1})$ is dual to the lattice $\ddot Q=\dot
Q(Q_1-2Q-Q_{-1}).$

In contrast to the relativistic case, the duality transformation is
irrelevant to integrability and cannot be used for the classification of the
lattices (\ref{tl}). For the first time this problem was solved by Yamilov
in the framework of the symmetry approach.

\begin{theorem}[Yamilov, \cite{Y93}] \label{th:tl}
The Toda type lattice (\ref{tl}) admits the higher symmetries iff
\begin{equation}\label{rf}
    r(p)=r_2p^2+r_1p+r_0, \quad f'=r_2f^2+R_1f+R_0.
\end{equation}
\end{theorem}

This result demonstrates that integrable lattices (\ref{tl}) can be obtained
from the integrable lattices of the relativistic Toda type by passing to the
limit $h=0$ in equations (\ref{rgh}).  More interesting link between this
two classes of equations can be established along the same arguments as in
the Section \ref{s:dtl}.

Let us denote iterations of transformation $T_l=T^{-1}_-T_+$ by the
superscript $l$ and let tilde in the formula (\ref{Tpm}) corresponds to the
value $l-1.$ Then formulae (\ref{Tpm}), (\ref{T}), (\ref{invT}) take form
\begin{align}
    & P_{-1}= h(y)p_{-1}+g(y),   \quad
     P^{-1}= h(y_1)p_1+g(y_1)   \label{PP}\\
    & p^1   = H(Y)P_{-1}+G(Y),   \quad
     p     = H(Y)P+G(Y)         \label{pp}
\end{align}
and therefore equations (\ref{rtl}), (\ref{RTL}) can be rewritten as a
coupled lattice of the Volterra type
\[
    \dot p=r(p)(P^{-1}-P_{-1}), \quad  \dot P=R(P)(p_1-p^1).
\]
Assuming
\begin{equation}\label{Pq}
    P=f(q_1-q^1), \quad f'=R(f)
\end{equation}
we obtain the Toda type lattice
\begin{equation}\label{tll}
    \ddot q= r(\dot q)(f(q^{-1}_1-q)-f(q-q^1_{-1}))
\end{equation}
for the variables situated along the line $l+n=\const.$  Since functions
$r,R$ were described in the Section \ref{s:rtl}, we immediately repeat the
Yamilov's result (\ref{rf}).

\subsection{Nonlinear superposition principle}\label{s:nsp}

Recall, that in the discrete case the 3-dimensional lattice generated by the
shifts $T_l,$ $T_m,$ $T_n$ contains 4 instances of the discrete relativistic
lattices and 3 instances of nonrelativistic ones (see Figure~3).  In the
continuous case the picture is more bare: the variables $q_{ln}$ form the
2-dimensional lattice containing 2 instances of the relativistics lattices,
1 instance of the Toda type lattices and 1 instance of the discrete Toda
type lattices, as shown on the Figure~4.

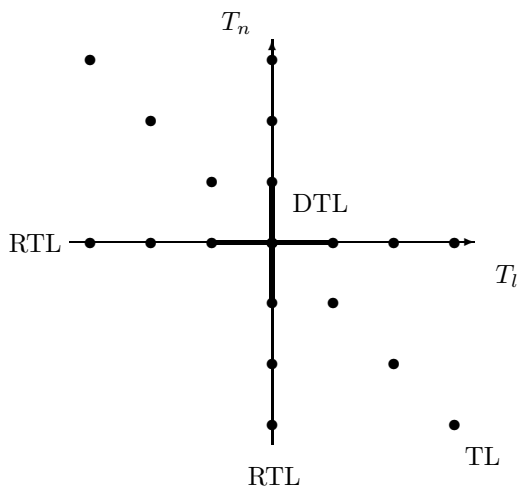
\begin{figure}[t]
\setlength{\unitlength}{0.07em}
\centering
\begin{picture}(250,240)(-130,-120) \small
 \put(-100,0){\vector(1,0){200}} \put(110,-20){$T_l$}
 \put(0,-100){\vector(0,1){200}} \put(-25,105){$T_n$}
 \multiput(-90,  0)( 30, 0){7}{\circle*{5}}  \put(-130, -5){RTL}
 \multiput(  0,-90)(  0,30){7}{\circle*{5}}  \put( -12,-120){RTL}
 \multiput( 90,-90)(-30,30){7}{\circle*{5}}  \put(  95,-110){TL}
 \linethickness{1.8pt}                       \put(  10,  15){DTL}
 \put(-30,0){\line(1,0){60}} \put(0,-30,0){\line(0,1){60}}
\end{picture}
\normalsize
\caption{Equations associated with the relativistic lattice.}
\end{figure}

The shift $T_n$ corresponds to the lattice (\ref{rtl}) and the diagonal
shift $T^{-1}_lT_n$ corresponds to the lattice (\ref{tll}), as was proved in
the previous section. The following Theorem demonstrates that the condition
of the commutativity of these shifts is equivalent to some discrete lattice
of the Toda type which can be interpreted as the nonlinear superposition
principle of the equations (\ref{rtl}) and (\ref{tll}).

\begin{theorem} \label{pr:dtl}
The variables $q,q^{\pm1},q_{\pm1}$ are related by equation of the form
\begin{equation}\label{dtl}
    (T_l-1)c(q^{-1}-q+\delta) + (T_n-1)c(q-q_{-1}) = 0.
\end{equation}
\end{theorem}
\begin{proof}
The function $\phi$ inverse to $A'$ satisfies equation $\phi'=R(\phi),$ that
is $f(y)=\phi(y+\eps).$  Therefore
\[
    C(Y) = A'(P)-y_1 = A'(f(q_1-q^1))-q_1+q = q-q^1+\eps,
\]
that is $Y=s(q-q^1+\eps)$ where $s$ is the inverse function to $C.$  It is
easy to prove that $s'$ satisfies the same equation (\ref{h}) as $h=c',$
that is $s(y)=c(y+\ti\eps)+\const.$

Next, eliminating $p$ from the formulae
\[
    Y=a'(p)-c(y_1), \quad Y^{-1}=a'(p)-c(y)
\]
brings to $Y^{-1}-Y=(T_n-1)c(y)$ and this completes the proof.
\end{proof}

It is clear that parameter $\delta$ can be set to zero by the proper choice
of the function $f$ in formula (\ref{Pq}) or, equivalently, by the shift
$q^l_n\to q^l_n+\delta l.$  Therefore, the terms in (\ref{dtl}) are
symmetric and the shifts $T_l$ and $T_n$ plays the equal roles, although
their origin was different.  This suggests that $T_l$ corresponds to some
relativistic lattice as well. Indeed, eliminating $P$ from the equation
$\dot p=r(p)(P^{-1}-P_{-1})$ by means of the relations (\ref{pp}) brings to
equation
\[
    \dot p=r(p)\left(-{p^1\over H(Y)}+{p^{-1}\over H(Y^{-1})}
                +{G(Y)\over H(Y)}-{G(Y^{-1})\over H(Y^{-1})}\right)
\]
and since $Y=c(q-q^1+\delta)$ we obtain a relativistic lattice again. The
coefficients of this equation coincide with $h,g$ up to the shift of
argument.


\resetfootnoterule

\subsection{Example: Heisenberg chain}\label{s:HC}

In the previous sections the discrete symmetries of integrable lattices were
studied.  Now we will briefly discuss the continuous symmetries. I restrict
myself by example of the equation (f$_0$)\footnote{These results were
obtained in collaboration with Professors A.B.~Shabat and A.P.~Veselov.}.
This choice is motivated by the link between (f$_0$) and the Heisenberg
chain which became very popular due to its applications in discrete geometry
\cite{Bob1,Bob2,DS}. Remind that this model reads
\begin{equation}\label{HC}
    s_t = as\times\left({s_1   \over 1+\<s,s_1\>}
                       +{s_{-1}\over 1+\<s,s_{-1}\>}\right)
          + b\left({s_1+s   \over 1+\<s,s_1\>}
                  -{s+s_{-1}\over 1+\<s,s_{-1}\>}\right)
\end{equation}
where $s\in\Real^3,$ $\<s,s\>=1,$ $\<,\>$ and $\times$ denote standard
scalar and vector products respectively, $a$ and $b$ are arbitrary
constants.  Up to the author knowledge this equation was introduced by
Sklyanin \cite{Skl} in the case $b=0$ and by Ragnisco and Santini \cite{RS}
in the general case.  The continuous limit of the Sklyanin lattice is the
Heisenberg model
\[
    s_t = s \times s_{xx}, \quad \<s,s\>=1.
\]
It was noticed that $r$-matrices for these models coincide and, more
generally, this property can be accepted as a definition of correct
discretization for a given continuous equation \cite{FT}.

In order to obtain (f$_0$) let us consider the complexification
$s\in\Real^3\to s\in\Complex^3$ and the stereographic projection
\begin{equation}\label{suv}
    s = S(u,v) = {1\over u-v}(1-uv, i+iuv, u+v).
\end{equation}
It is convenient to represent the flow (\ref{HC}) for arbitrary set of
parameters $a,b$ as a linear combination of the flows corresponding to the
sets $a=i,b=\pm 1.$  These flows are given by the following formulae in
terms of the variables $u,v:$
\begin{gather}
    u_{t_+}= {(u_1-u)(u-v)   \over u_1-v},\quad
    v_{t_+}= {(u-v)(v-v_{-1})\over u-v_{-1}},   \label{uvt+} \\
    u_{t_-}= {(u_{-1}-u)(u-v)\over u_{-1}-v},\quad
    v_{t_-}= {(u-v)(v-v_1)   \over u-v_1}.      \label{uvt-}
\end{gather}
The lattice (\ref{uvt+}) appeared in \cite{SY87,SY90} for the first time.

Next, notice that elimination of $P$ in virtue of (\ref{Pq}) brings the
formulae (\ref{PP}) to the form
\[
    \dot q = {f(q_1-q^1) - g(q_1-q) \over h(q_1-q)}, \quad
    \dot q^1 = {f(q-q^1_{-1}) - g(q^1-q^1_{-1}) \over h(q^1-q^1_{-1})}.
\]
It is easy to see that this system corresponding to the case (f$_0$)
($f(y)=g(y)=-1/y,$ $h(y)=1/y^2$) coincide with (\ref{uvt+}) if we assume
$q=u,$ $q^1=v$ and $t=t_+.$  Since the lattice (\ref{uvt-}) is obtained by
reflection $n\to -n$ we immediately come to the following statement.

\begin{prop}
Both variables $u$ and $v$ satisfy the lattices
\[
    q_{t_\pm t_\pm} = q^2_{t_\pm}\left(
                        {q_{\pm1,t_\pm} \over (q_{\pm1}-q)^2}
                      - {q_{\mp1,t_\pm} \over (q-q_{\mp1})^2}
                                   - {1 \over  q_{\pm1}-q}
                                   + {1 \over  q-q_{\mp1}} \right)
\]
in virtue of the equations (\ref{uvt+}), (\ref{uvt-}).
\end{prop}

The lattices (\ref{uvt+}), (\ref{uvt-}) are Hamiltonian.  Their Hamiltonians
are
\[
    H_+= \sum_n\log{u_1-v\over u-v}, \quad H_-= \sum_n\log{u-v_1\over u-v}
\]
respectively and the Poisson brackets are of the form
\begin{equation}\label{pois}
    \{u_m,v_n\}=(u_n-v_n)^2\delta_{mn}, \quad  \{u_m,u_n\}=\{v_m,v_n\}=0.
\end{equation}

The next proposition can be proved by straightforward calculations.  It
demonstrates that the lattices (\ref{uvt+}), (\ref{uvt-}) are the symmetries
of each other and the shift $(u_n,v_n)\to(u_{n+1},v_{n+1})$ defines the
B\"acklund transformation for some hyperbolic system.

\begin{prop}
The Hamiltonians $H_+$ and $H_-$ are in involution, the vector fields
$\pa_{t_+}$ and $\pa_{t_-}$ commute and the variables $u_n,v_n$ satisfy the
system
\begin{equation}\label{uvtt}
    u_{t_+t_-}= {2u_{t_+}u_{t_-}\over u-v} - u_{t_+} - u_{t_-}, \quad
    v_{t_+t_-}= {2v_{t_+}v_{t_-}\over v-u} + v_{t_+} + v_{t_-}.
\end{equation}
In terms of the vector (\ref{suv}) this system reads
\[
    s_{t_+t_-} + \<s_{t_+},s_{t_-}\>s + is\times(s_{t_+}+s_{t_-}) = 0,
    \quad \<s,s\>=1.
\]
\end{prop}

Commutativity of the flows $\pa_{t_+},$ $\pa_{t_-}$ allows to construct
compatible zero curvature representations
\[
    W_{t_+} = U^+_1W-WU^+, \quad  W_{t_-} = U^-_1W-WU^-,
\]
so that $\tr W_N\dots W_1$ generates the common first integrals of the both
lattices under the periodic boundary conditions $u_N=u_0,$ $v_N=v_0.$  The
matrices $W,U^\pm$ are given by the formulae
\[
    W = \la I + P(u,v), \quad
    U^+ = {1\over2\la}(P(u,v_{-1})-P(v_{-1},u)),
\]\[
    U^- = {1\over2(\la+1)}(P(v,u_{-1})-P(u_{-1},v))
\]
where $P$ denotes the projector
\[
    P(u,v)= {1\over u-v}
              \left(\ba{cc}
                -v & uv \\
                -1 & u
              \ea\right).
\]

The Poisson structure (\ref{pois}) can be written in the $r$-matrix form
\[
    \{W_m(\la)\ot_,W_n(\mu)\} = \delta_{mn}[r,W_m(\la)\otimes W_n(\mu)]
\]
with the same $r$-matrix as for the Heisenberg model:
\[
    r = {1\over\la-\mu}
          \left(\ba{cccc}
            1 & 0 & 0 & 0 \\
            0 & 0 & 1 & 0 \\
            0 & 1 & 0 & 0 \\
            0 & 0 & 0 & 1
          \ea\right).
\]

Equation (\ref{uvtt}) gives an example of hyperbolic symmetry.  Any lattice
(\ref{rtl}) of the relativistic Toda type admitting duality transformations
possesses also evolution symmetries, the simplest representative is given in
the following Theorem.

\begin{theorem}[\cite{AS2}] \label{th:sym}
The Lagrangian (\ref{L}) admits the variational symmetry of the form
\begin{equation}\label{qtau}
    q_\tau= r(p)(h(y_1)p_1+h(y)p_{-1}+g(y_1)+g(y))+R_1p^2
\end{equation}
if and only if the coefficients $r,g,h$ satisfy the system (\ref{rgh}).
\end{theorem}

Obviously, one can rewrite this symmetry as the partial differential
equation if use the lattice (\ref{rtl}) itself and assume $q=u,$ $q_1=v:$
\begin{align*}
  u_\tau + u_{tt} &= 2r(u_t)(h(v-u)v_t+g(v-u))+R_1u^2_t, \\
  v_\tau - v_{tt} &= 2r(v_t)(h(v-u)u_t+g(v-u))+R_1v^2_t.
\end{align*}
Returning to the system (\ref{uvt+}) one obtains the symmetry ($t=t_+$)
\begin{equation} \label{uvtau}
\begin{split}
  iu_\tau - u_{tt} &=
   2u^2_t\left({v_t\over(u-v)^2}-{1\over u-v}\right), \\
  iv_\tau + v_{tt} &=
   2v^2_t\left({u_t\over(u-v)^2}-{1\over u-v}\right)
 \end{split}
\end{equation}
which is equivalent, in the geometrical terms (\ref{suv}), to the modified
Heisenberg model \cite{MS}
\begin{equation}\label{stau}
    s_\tau = s\times s_{tt} - {i\over2}s_t\<s_t,s_t\>, \quad \<s,s\>=1.
\end{equation}
So we have proved that the flows defined by the lattice (\ref{HC}) at
$a=i,b=1$ and equation (\ref{stau}) commute.  One can reformulate this
result as follows:

\begin{prop}
The shift $s\to s_1$ in the Heisenberg chain (\ref{HC}) for $a=i,b=1$
defines the $t$-part of the B\"acklund transformation for equation
(\ref{stau}).
\end{prop}

Notice, that according to (\ref{uvt+}) this shift is equivalent to solving
of Riccati equation on the variable $v_1:$
\[
    u_t= {(u_1-u)(u-v)\over u_1-v},\quad
    v_{1,t}= {(u_1-v_1)(v_1-v)\over u_1-v}.
\]

Another choice of dependent variables allows to rewrite (\ref{uvtau}) in the
form
\[
    iu_{1,\tau} =  u_{1,tt} - {2u^2_{1,t}\over u_1-v}, \quad
    iv_\tau = -v_{tt} - {2v^2_t\over u_1-v}.
\]
This system is equivalent to the Heisenberg model
$\sigma_\tau=\sigma\times\sigma_{tt},$ $\<\sigma,\sigma\>=1$ in terms of the
vector $\sigma=S(u_1,v),$ which is related with the vectors (\ref{suv}) by
the formula
\[
    \sigma = {s_1+s-is_1\times s \over 1+\<s_1,s\>}.
\]

\section{Multifield examples}\label{s:dmult}

In the previous sections we assumed that the variable $q$ was scalar
($q\in\Field,$ $\Field=\Real,\Complex$).  The brief analysis proves that the
Theorems \ref{th:ddt} and \ref{th:dt} remain valid for the vector case
$q\in\Field^N$ as well.  Thus, the multifield generalizations can be
obtained by finding of the mappings $T:\Field^{2N}\to\Field^{2N}$ with the
special structure described in these Theorems.  This problem is much more
difficult than in the scalar case and the classification of such mappings is
far from completeness.

However it is not difficult to find some particular examples.  The most
simple are the generalizations of the discrete Heisenberg equation (A) and
the Heisenberg lattice (f$_0$) related to the Jordan triple systems. It was
Svinolupov who recognized the role of the Jordan algebraic structures in the
theory of integrable systems for the first time.  I provide only necessary
information about the Jordan triple systems, the interested reader can find
more in \cite{Sv,SvY91,SvY94,HSY,ASY} and references therein.

\subsection{Jordan triple systems}

The ternary algebra $J$ with multiplication $\{\}: J^3\to J$ is called
Jordan triple system if the following identities hold
\begin{gather}
 \{abc\} = \{cba\},                                          \label{J3}\\
 \{ab\{cde\}\}-\{cd\{abe\}\} = \{\{cba\}de\}-\{c\{bad\}e\}.  \label{J5}
\end{gather}
Some consequences are:
\begin{equation}\label{id}
\begin{split}
 &\{ab\{aca\}\} = \{a\{bac\}a\} = \{\{aba\}ca\}, \quad
    \{a\{bab\}c\} = \{\{aba\}bc\},                      \\
 &  2\{\{abc\}bd\} = \{a\{bcb\}d\} + \{c\{bab\}d\}.
\end{split}
\end{equation}
Calculations are simplified by use of the linear operators $L_{ab}, M_{ab},
M_a: J\to J$ defined for arbitrary elements $a,b\in J$ as follows:
\[
    L_{ab}(c)=\{abc\},\quad M_{ab}(c)=\{acb\}, \quad M_a(c)=M_{aa}(c)=\{aca\}.
\]
Notice, that identity (\ref{J5}) is equivalent to
\begin{equation}\label{LLid}
    [L_{ab},L_{cd}] = L_{\{cba\}d} - L_{c\{bad\}}.
\end{equation}

We are especially interested in rational expressions.  They are build from
the inverse elements which are defined as $a^{-1}=M^{-1}_a(a).$  In some
Jordan triple systems the operator $M_a$ is degenerate.  In such cases the
notion of inverse element can be partly substituted by the notion of a
deformation vector which is the solution of the system $\pa b/\pa a=-M_b$
\cite{SS95}.  However, for sake of simplicity, we shall consider only the
cases when $\det M_a\ne0$ almost everywhere in $J.$   The Jordan triple
system with this property are called Jordan triple system with invertible
elements.

The following Lemma demonstrates that some properties of $a^{-1}$ are the
same as in a ring.

\begin{lemma}\label{l:inv}
Let the operator $M_a$ be invertible then
\begin{equation}\label{inva}
    M_{a^{-1}} = M^{-1}_a, \quad (a^{-1})^{-1} = a, \quad (a^{-1})_x =
    -M^{-1}_a(a_x)
\end{equation}
and the following rules of cancellation of $a$ and $a^{-1}$ are valid
\begin{equation}\label{inva'}
\begin{split}
  & \{aa^{-1}b\}=b, \quad
    \{a^{-1}\{acb\}a^{-1}\}=\{cba^{-1}\},   \\
  & \{a\{a^{-1}ca^{-1}\}b\}=\{ca^{-1}b\}, \quad
    \{ba\{a^{-1}ca^{-1}\}\}=\{bca^{-1}\}.
\end{split}
\end{equation}
\end{lemma}
\begin{proof}
Let $b\in J,$ $c=M_a(b),$ $d=M_{a^{-1}}(c).$  One obtains using (\ref{id})
\begin{align*}
    M_a(d)&= \{a\{a^{-1}ca^{-1}\}a\}
           =  2\{aa^{-1}\{aa^{-1}c\}\} - \{a\{a^{-1}aa^{-1}\}c\} \\
          &= 2\{aa^{-1}\{aa^{-1}\{aba\}\}\} - \{\{aa^{-1}a\}a^{-1}c\}
           =  \{aa^{-1}\{aba\}\} \\
          &= \{aba\} = M_a(b),
\end{align*}
and therefore $b=d.$  Arbitrariness of $b$ implies $M_{a^{-1}}M_a=I.$

Further on, $(a^{-1})^{-1}= M^{-1}_{a^{-1}}(a^{-1})=M_a(a^{-1})=a.$

For arbitrary $b$
\[
    \{aa^{-1}b\} = \{aa^{-1}\{aM^{-1}_a(b)a\}\} =
                \{\{aa^{-1}a\}M^{-1}_a(b)a\} = \{aM^{-1}_a(b)a\} = b.
\]
Taking this into account when differentiating relation $a=\{aa^{-1}a\}$ one
proves the last formula in (\ref{inva}).

In virtue of (\ref{LLid}) and relation $L_{aa^{-1}}=I$ which is already
proved one obtains
\[
    L_{M_{a^{-1}}(c)a} = L_{a^{-1}\{ca^{-1}a\}} + [L_{a^{-1}c},L_{a^{-1}a}] =
    L_{a^{-1}c} + [L_{a^{-1}c},I] = L_{a^{-1}c},
\]
and the last formula in (\ref{inva'}) is proved.  The next to the last is
equivalent (use operator notation).  Finally, (\ref{J5}) implies
\[
    \{a^{-1}\{acb\}a^{-1}\} = 2\{a^{-1}bc\} - L_{M_{a^{-1}}(b)a}(c) =
    \{a^{-1}bc\}.    \quad\qed
\]
\renewcommand{\qed}{}\end{proof}

Now we can prove the formula (``harmonic mean'')
\begin{equation}\label{harm}
    (a^{-1}+b^{-1})^{-1} = \{a(a+b)^{-1}b\}.
\end{equation}
Let us denote $a+b=c^{-1},$ then
\begin{align*}
   M_{(a^{-1}+b^{-1})}(\{acb\})
    = {}& \{a^{-1}\{acb\}a^{-1}\} + \{b^{-1}\{acb\}b^{-1}\} \\
      & + 2\{a^{-1}\{ac(c^{-1}-a)\}b^{-1}\}  \\
    = {}& \{cba^{-1}\} - \{b^{-1}ac\} + 2b^{-1}.
\end{align*}
Symmetrization on $a$ and $b$ gives (\ref{harm}).

Analogue of the Killing form in the Jordan triple system is the scalar product
$\<a,b\>=\tr L_{ab}.$  Relation (\ref{LLid}) implies the invariance property
\begin{equation}\label{scid}
    \<\{abc\},d\>= \<a,\{bcd\}\>
\end{equation}
of this product.  Further on we assume that it is also symmetric and
nondegenerate.  Notice that if the element $a$ is invertible then the
equalities $L_{ab}=M_aM_{a^{-1}b},$ $L_{ba}=M_{a^{-1}b}M_a$ imply
$\<a,b\>=\<b,a\>.$ Hence, in virtue of continuity, in the Jordan triple
systems with invertible elements the symmetry property always holds.

Examples below together with the reductions $a=\pm a^\tau$ of the Examples 1
and 2 (with $M=N$) exhaust all simple Jordan triple systems aside from two
exceptional ones.

\paragraph{Example 1.}
The linear space $J$ of $N\times N$ matrices becomes the Jordan triple
system with respect to the triple product defined by means of the standard
matrix multiplication as follows
\[
    \{abc\} = {1\over2}(abc+cba).
\]
The operator $M_a$ is invertible iff $\det a\ne0,$ that is almost
everywhere. The element $a^{-1}$ coincides with inverse matrix.  The
subspaces of symmetric and skewsymmetric matrices are Jordan triple systems
as well.  However, in the Jordan triple system of the skewsymmetric matrices
of odd order the operator $M_a$ is degenerate for all $a.$  The scalar
product is $\<a,b\>=\tr ab.$

\paragraph{Example 2.}
Previous example admits generalisation for $N\times M$ matrices:
\[
    \{abc\} = {1\over2}(ab^\tau c + cb^\tau a),
\]
where $^\tau$ denotes transposition.  In particular, if $M=1$ then $J$ turns
into the $N$-dimensional vector space with multiplication
\[
    \{abc\} = {1\over2}(\<a,b\>c + \<c,b\>a)
\]
where $\<,\>$ denotes the standard scalar product.  However, the operator
$M_a$ is not invertible for $M\ne N$.

\paragraph{Example 3.}
More interesting triple product in the $N$-dimensional vector space is given
by
\[
    \{abc\} = \<a,b\>c + \<c,b\>a - \<a,c\>b.
\]
The scalar product in $J$ coincides with the standard one.  Operator $M_a,$
its inverse and vector $a^{-1}$ are defined by formulae
\[
    M_a(b)=2\<a,b\>a-\<a,a\>b, \quad M^{-1}_a=\<a,a\>^{-2}M_a, \quad
    a^{-1}=\<a,a\>^{-1}a.
\]

\subsection{Jordan analogues of Heisenberg equations}

The lattice (\ref{drtl}), (A) of discrete relativistic Toda type admits
literal generalization
\begin{equation}\label{JA}
    \mu(T_m-1)x^{-1} + \nu(T_n-1)y^{-1} +\la(T_mT_n-1)z^{-1} = 0,
    \quad \la+\mu+\nu=0
\end{equation}
for arbitrary Jordan triple system with invertible elements.  Indeed, using
relation (\ref{harm}) one can prove that the mapping $T:J^2\to J^2$ given by
the formulae
\[
    X=  \nu y^{-1} + \la(x+y)^{-1},\quad
    Y= -\mu x^{-1} - \la(x+y)^{-1}
\]
coincide with its inverse and therefore the duality transformations
(\ref{dTpm}) map equation (\ref{JA}) into itself.  The nonrelativistic
analogue of this equation corresponding to the set of parameters $\mu=1,$
$\nu=-1,$ $\la=0$ can be obtained along the same scheme as in the Section
\ref{s:dtl}.

Equation (\ref{JA}) is the Euler equation for the Lagrangian
\[
    {\cal L}=\sum_{m,n}(\mu\log\det M_x+\nu\log\det M_y+\la\log\det M_z).
\]
In order to prove this, note that if $f(u)=\12\log\det M_u$ then, in virtue
of Lemma \ref{l:inv},
\[
  \<{\pa f\over\pa u},v\> = {d\over d\eps}f(u+\eps v)|_{\eps=0}
  = \tr(M^{-1}_uM_{uv})= \tr L_{u^{-1}v} = \<u^{-1},v\>
\]
and therefore $\pa f/\pa u=u^{-1}.$

Analogously, the lattice (f$_0$) of the relativistic Toda type admits
generalization
\[
    \dot y= p-p_{-1}, \quad
    \dot p= M_p(M^{-1}_{y_1}(p_1) - M^{-1}_y(p_{-1}) - y^{-1}_1 + y^{-1})
\]
corresponding to the Lagrangian
\[
    {\cal L}= \int dt\sum_n(\12\log\det M_p-\<p,y^{-1}\>-\12\log\det M_y).
\]
Indeed, if $f(u)=\<u^{-1},a\>$ then
\[
    {d\over d\eps}f(u+\eps v)|_{\eps=0} = -\<M^{-1}_u(v),a\> =
    -\<M^{-1}_u(a),v\>
\]
and therefore $\pa f/\pa u= -M^{-1}_u(a).$

The duality transformations (\ref{Tpm}) are defined by the mapping $T$ of
the form
\[
    Y = p^{-1}-y^{-1},\quad P = y^{-1}-M^{-1}_y(p)
\]
which is involutive.  Applying the scheme of the Section \ref{s:tl} one
obtains the Toda type lattice
\[
    \ddot q= -M_{\dot q}((q_1-q)^{-1} - (q-q_{-1})^{-1})
\]
which is equivalent to the Jordan Volterra lattice \cite{ASY}
\[
    \dot p= M_p(P-P_1),\quad  \dot P= M_P(p_{-1}-p).
\]

\paragraph{Acknowledgements.}  Author thanks Professors B.A.~Kupershmidt and
A.B.~Shabat for useful discussions.  This work was supported by the grant \#
99-01-00431 of the Russian Foundation for Basic Research.


\label{adler_lp}

\end{document}